\documentclass{emulateapj}

\usepackage{amsmath}
\usepackage{color}
\usepackage{graphicx}
\usepackage{multirow}
\usepackage{lipsum}

\begin{document}
\title{X-ray Spectra from Plasmas with High-Energy Electrons: $\kappa$-distributions and \MakeTextLowercase{e}$^{-}$-\MakeTextLowercase{e}$^{-}$  Bremsstrahlung}
\author{Xiaohong Cui\altaffilmark{1,2}, Adam R. Foster\altaffilmark{1}, Takayuki Yuasa\altaffilmark{3}, Randall K. Smith\altaffilmark{1}}
\altaffiltext{1}{Smithsonian Astrophysical Observatory, 60 Garden Street, Cambridge, MA 02138, USA}
\altaffiltext{2}{National Astronomical Observatories, Chinese Academy of Sciences, 20A Datun Road, Chaoyang District, Beijing, China} 
\altaffiltext{3}{yuasatakayuki@gmail.com}

%
%

\begin{abstract}
Shocks, turbulence and winds all influence the electron velocity distribution in hot plasmas, exciting lower-energy electrons and generating a high-energy (typically power-law) tail. This effect, typically described as a $\kappa$ distribution can affect both the line and continuum X-ray spectrum emitted by the plasma. Hahn \& Savin (2015) proposed a ``Maxwellian decomposition'' to generate the rate coefficients of $\kappa$ distributions. Using their method and the AtomDB atomic database, we have developed a general model to calculate the emission from a plasma with a $\kappa$ distribution. We compare our $\kappa$ results for the charge state distribution and spectra of oxygen to those from KAPPA package with the ion data available within the CHIANTI atomic database. Sufficiently energetic electrons, created either in a $\kappa$ distribution or merely a very hot Maxwellian plasma, can also emit via electron-electron (e-e) bremsstrahlung, a process not previously included in AtomDB. We have added this process to AtomDB and apply it to calculate the temperature gradients, as well as the total spectra from the post-shock regions of an accreting magnetic cataclysmic variable (CV). We find the contribution of e-e bremsstrahlung to the total spectra exceeds 10\% at KT$\sim 100$ keV, with the total emissivity in the post-shock accretion stream differing by more than 10\% at energies above 60 keV.
\end{abstract}


\section{Introduction}
Typically, physical properties of a collisional plasma are derived under the
assumption that the electrons are in a local equilibrium Maxwell-Boltzmann distribution (hereafter referred as
Maxwellian distribution) characterized by a single temperature.  The thermalization timescale for electron-electron collisions to create a Maxwellian energy distribution in the plasma is generally short compared with other processes (Summers et al. 2006). Thus it is usually the case that the free electrons have close to a Maxwellian distribution. However, in real astrophysical plasmas, some forces, e.g., shocks, turbulence, or wave-particle interactions, can drive the plasma to a relatively long-lasting non-Maxwellian distribution.

The first direct identification of a non-Maxwellian electron distribution was found by Vasyliunas (1968) in the Earth's magnetosphere. The electron
distribution driven by the electromagnetic forces could be fitted using a near-Maxwellian core and a power-law tail in high energy called a $\kappa$ distribution. Since then, $\kappa$ distributions have been utilized in numerous studies of solar physics (e.g. Collier et al. 1996; Livadiotis \& McComas 2010; Le Chat et al. 2011; Dud\'ik et al. 2015, 2017; Jeffrey et al. 2016, 2017; Dzif\v{c}\'akov\'a et al. 2018; Livadiotis et al. 2018) and planetary magnetospheres (e.g. Christon, 1987; Mauk et al. 2004; Diayllynat et al. 2009; Henning et al. 2011; Carbary et al. 2014; Barbosa et al. 2016; Pollock et al. 2017).

The $\kappa$ distribution has been the subject of considerable interest in the solar system as well as in other astrophysical plasmas, e.g., supernova remnants, planetary nebulae, and galaxy clusters. In solar plasmas, $\kappa$ distributions can arise when the plasma is being continually pumped by a non-thermal energy input (e.g. a shock) or by energy transport from other places. Raymond et al. (2010) found a spectrum of a slit position in the northeast of Tycho's supernova remnant could be well fitted with $\kappa$ distribution, implying a shock speed of 2800 km s$^{-1}$. A $\kappa$
electron energy distribution was adopted to give consistent temperature measurements and metalicity estimates in HII regions and planetary nebulae (Nicholls et al. 2012). This distribution was also considered in clusters of galaxies (Petrosian \& East 2008; Kaastra et al. 2009), where the processes involving strong turbulence or plasma acceleration can produce a non-thermal electron distribution. Reviews on $\kappa$-distributions and their applications in astrophysical plasma can be found in Pierrard \& Lazar (2010), Bykov et al. (2013) and Livadiotis (2018).

Diagnostics of $\kappa$-distributions in the solar corona were attempted using extreme ultraviolet lines (Dzif\v{c}\'akov\'a \& Kulinov\'a 2010; Mackovjak et al. 2013) for the plasma temperature, electron density, and $\kappa$ value. A diagnostic of non-thermal particle heating in small coronal heating events or nanoflares (Testa et al. 2014) was also detected in IRIS observations (De Pontieu et al. 2014). A thermal Maxwellian component plus a power-law tail in the high energy range can explain some of observed X-ray flare line spectra (Lin et al. 2002; Ka\v{s}parov\'a \& Karlick\'y 2009; Dzif\v{c}\'akov\'a et al. 2011).

The $\kappa$-distribution can arise naturally from theory. A plasma wave field can induce fluctuation which causes the non-Coulombic diffusion in velocity space. The diffusion is proportional to the square of the particle velocity in plasma and leads to a $\kappa$-distributed electrons (Hasegawa et al. 1985; Hasegawa \& Sato 1989, note that the $\kappa$ definition used in these papers is equivalent to this $\kappa - 1$).
Tsallis statistical mechanics has been used to describe and analyze complex systems out of equilibrium (Leubner 2002; Collier 2004; Livadiotis \& McComas 2009, 2010, 2013). In these systems there are correlations induced by any long-range interactions between the particles and the specific formulation of the $\kappa$-distribution can mathematically model the correlations between particles. The equality between the kinetic and thermodynamic temperatures produces a well-defined ``temperature'' for systems out of thermal equilibrium. The physical meaning of the kappa index is then connected with the correlation of particles in the systems.

Based on the CHIANTI database, Dzif\v{c}\'{a}kov\'{a} et al. (2015) developed the KAPPA package to calculate the synthesis of optically thin spectra for the non-Maxwellian $\kappa$ distributions. Both ionization and recombination rates with the ionization equilibria were reverse-engineered from the Maxwellian recombination rates and tabulated in the KAPPA package for a range of $\kappa$ values (Dzif\v{c}\'akov\'a \& Dud\'ik 2013; Wannawichian et al. 2003; Dzif\v{c}\'akov\'a 1992). The ``reverse-engineering'' approach was to fit the rate coefficient for Maxwellian distributions, extract an approximate cross section and then reconvolve with the desired $\kappa$ distribution. Another method, ``Maxwellian decomposition'' was proposed to generate the rate coefficients of $\kappa$ distributions by summing the appropriately weighted Maxwell-Boltzmann rate coefficients (Ko et al. 1996, Kaastra et al. 2009, Hahn \& Savin 2015).

The power law tail of hot electrons in a $\kappa$ distribution emphasizes the importance of hot electrons, and emission processes which are relevant to them. In particular, the electron-electron (e-e) thermal bremsstrahlung occurs when an electron moves through another electron. The acceleration by the Coulomb force to both electrons are equal but the direction is opposite. The wave train emitted by one electron
destructively interferes with that emitted by the other. This leading to very low emission of e-e thermal bremsstrahlung at non-relativistic temperatures. This simultaneity disappears in relativistic conditions and hence e-e bremsstrahlung becomes important. It is calculated from the integral of particle velocity by averaging the cross section over the Maxwellian distribution was calculated in previous works, e.g. Maxon \& Corman (19670), Haug (1975a,b), Stepney \& Guilbert (1983). At temperature T$\sim10^8$ K, the integral is less than 7\% than that of electron-proton (e-p) bremsstrahlung at 30 keV and increases with the photon energy. At relativistic temperatures, both electrons in e-e bremsstrahlung can be thought of as radiating, giving a spectrum about twice as large as for e-p bremsstrahlung, where only one of the interacting particles radiates (Svensson 1982). The e-e bremsstralung emissivity was therefore to be non-negligible compared to the electron-ion bremsstrahlung at temperatures above 10$^8$ K for Maxwellian plasmas. In $\kappa$ plasmas, this contribution will be even higher.

The AtomDB atomic database (Smith et al. 2001, Foster et al. 2012) collects the astrophysical plasma emission data relevant to X-ray emission from collisionally ionized, optical thin astrophysical plasmas with temperature $10^4 \leq$T$_e \leq10^9$ K. The atomic data stored includes ionization and recombination rates, energy levels, wavelengths, radiative transitions and electron impact collision strengths. This data is then coupled with the Astrophysical Plasma Emission Code (APEC, Smith 2001) to produce emissivities for optically thin, collisionally ionized plasma with Maxwellian electron distributions. We have used AtomDB 3.0.9, which includes a set of non-equilibrium ionization (NEI) emissivity files, designed to allow modeling of an out-of-equilibrium plasma. We have also used new version of PyAtomDB to access the database and generate our spectra. The models we have used are compatible with XSPEC and are available online\footnote{https://www.atomdb.org/kappa}.

We use these tools to apply the Maxwell decomposition method to get the $\kappa$ line and continuum spectra, focusing here on oxygen as an exemplar. We also calculate the bremsstrahlung emission and dielectronic (DR) and radiative recombination (RR) rate coefficients for a $\kappa$ distribution of electrons directly to compare the accuracy of the decomposition approach with an exact model\footnote{https://github.com/AtomDB/kappa}. In this paper, we describe the processes (Sect. 2) and method (Sect. 3) to produce the X-ray optically thin astrophysical spectra arising from collisional excitation by electrons with a $\kappa$-distribution and that with e-e bremsstrahlung emissions. Spectra with $\kappa$-distributed electrons and e-e bremsstrahlung, as well as the application of the model to the numerical post-shock accretion region (PSR) model of a white dwarf are presented in Sect. 4. Conclusions are given in Sect. 5.

\section{Processes}
\label{proce}
The isotropic $\kappa$-distribution of the electron energies is given by (e.g., Olbert et al. 1967; Owocki \& Scudder 1983),
\begin{equation}
\label{eq:Fk}
\resizebox{0.5\textwidth}{!}{$f_{\kappa}(E; \kappa,T)=A_{\kappa}\frac{2}{\sqrt{\pi}}(\frac{1}{k_{\rm B}T})^{3/2}\sqrt{E}[1+\frac{E}{(\kappa-3/2)k_{\rm B}T}]^{-(\kappa+1)}$},
\end{equation}
with
$A_\kappa=\frac{\Gamma(\kappa+1)}{\Gamma(\kappa-1/2)(\kappa-3/2)^{3/2}}$
and Gamma function $\Gamma$. $k_{\rm B}$ is the Boltzmann constant and $T$ is the temperature. The parameter $\kappa$ is in the range of (3/2,$+\infty$) with $\kappa \rightarrow +\infty$ being a Maxwellian distribution and $\kappa \rightarrow 3/2$ corresponding to larger departure from thermal distribution (Hasegawa et al. 1985; Hasegawa \& Sato 1989;  Summers \& Thorne 1991; Mace \& Hellberg 1995; Lee \& Jung 2019). From Eq. (\ref{eq:Fk}), the $\kappa$-distribution can be shown to behave as a Maxwellian with temperature $T_{\rm M}=\frac{\kappa-3/2}{\kappa+1}T$ in the low-energy limit (Meyer-Vernet et al. 1995; Livadiotis \& McComas 2009) and it has a power law tail proportional to $E^{-(\kappa+1)}$ for high energies.

The collisional rates in a collisionally ionized plasma are derived by integrating the relevant cross sections over the electron distribution. If we can decompose an electron distribution into a linear combination of Maxwellian components, the rate coefficient for any process is simply the sum of the rates for these individual Maxwellian components.  This Maxwellian decomposition method has regularly been used to approximate $\kappa$-distributions with a sum of Maxwellians (Ko et al. 1996, Kaastra et al. 2009, Hahn \& Savin 2015). Hahn \& Savin (2015) approximated the $\kappa$-distribution for a wide range of values using a formula
\begin{equation}
\label{eq:fkappa}
f_\kappa(E;\kappa,T_\kappa) = \displaystyle{\sum_{j}} c_j f(E;a_jT_\kappa),
\end{equation}
where $f(E;a_jT_\kappa)$ is the Maxwellian energy distribution at a temperature $T_{\rm M} = a_jT_\kappa$. They obtained and tabulated the values of best fit parameters $a_j$ and $c_j$ for $\kappa$ values from 1.7 to 100.

For Maxwellian electron distributions, all collisional rates (recombination, ionization, excitation) are tabulated in AtomDB between $T_{\rm M}=10^4-10^9$ K. In this paper, we use oxygen as an example for applying this deconvolution method. We create ionization and recombination rates for $\kappa$ plasmas from the AtomDB Maxwellian rates, use these to obtain the ion fraction, then calculate the emissivity per ion using the same method, and then multiply the ion fraction by the kappa emissivities to get non-Maxwellian emission.

The recombination rates can also be calculated directly by integrating the radiative recombination cross section over a non-Maxwellian electron distribution. As comparisons, we give such kind of calculations for the rates of radiative recombination and dielectronic recombination. For the bremsstrahlung emission with $\kappa$-distributions of electrons, we integrate the free-free cross section (gaunt factor) and $\kappa$-distribution to get the energy spectrum.

The emitted spectrum can be separated into lines and continuum emission. In our framework, lines due to collisional excitation can only be calculated using the Maxwellian decomposition method, while those from DR can be obtained either by analytical calculation or by Maxwellian decomposition. We consider three radiative processes for the continuum emission: thermal bremsstrahlung (i.e. free-free emission), radiative recombination continuum (RRC, i.e. free-bound emission), and two-photon radiation (i.e. bound-bound emission). The most important continuum emission is the thermal bremsstrahlung for optically-thin thermal plasma in a collisional ionization equilibrium (CIE). Following we will present the calculations of the bremsstrahlung, RRC, DR processes, and the CSD from collisional rate coefficients.

\subsection{Bremsstrahlung} 
The bremsstrahlung emission, also called free-free radiation, occurs when an unbound electron moves through the electric field of a nucleus or another electron and is accelerated by the Coulomb force. This process results in continuum radiation. The non-relativistic free-free Gaunt factor was first discussed with the detailed numerical computations in the work of Karzas \& Latter (1961) and then has been the subject of a series of publications (e.g. Sutherland 1998; Nozawa et al. 1998; Itoh et al. 2000; van Hoof et al. 2014, 2015; de Avillez \& Breitschwerdt 2015) that followed that work.

Assuming the initial energy of electron is $E_i$, the final energy is $E_f=E_i-h \nu$ after being accelerated by a nucleus with atomic number $Z$ and emitting a photon with energy $h \nu$. The free-free Gaunt factor is (Karzas \& Latter 1961)
\begin{equation}
\label{eq:ff_gaunt}
\begin{split}
g_{ff}= & \frac{2\sqrt{3}}{\pi \eta_i \eta_f}[(\eta_i^2+\eta_f^2+2\eta_i^2\eta_f^2)I_0\\
&-2\eta_i\eta_f\sqrt{(1+\eta_i^2)(1+\eta_f^2)}I_1]I_0 , 
\end{split}
\end{equation}
where $\eta_i^2 =\frac{Z^2R_y}{E_i}$, $\eta_f =\frac{Z^2R_y}{E_f} $, and Ry is the
infinite-mass Rydberg unit of energy (13.6eV). The functions $I_{\rm l}$ and $I_{\rm 0}$ were expressed to be a series solutions of certain differential equations related with hypergeometric functions in the work of Karzas \& Latter (1961). After taking into account the effect of the electron degeneracy, Itoh et al. (1985, 1990, 2000) and Nozawa et al. (1998) extended Karzas \& Latter's calculations to the relativistic regime and presented the accurate analytic formulae of relativistic cross section of free-free opacity.

If the velocity distribution of the electron is $f(\upsilon, T)$  at temperature $T$, the energy spectrum of bremsstrahlung emission at temperature $T$ from these electrons is $\varepsilon_{ff}(\nu,T)=4 \pi \int ^{\infty}_{x} f(\upsilon, T) \omega(\upsilon)d \upsilon$ (e.g., Kwok 2007; Phillips et al. 2008), where $\omega(\upsilon)=\frac{8\pi e^6}{3\sqrt{3}c^3m_e^2 \upsilon}n_en_iZ^2g_{ff}(\nu, \upsilon)$ is the power radiated in unit volume per solid angle per unit frequency, $e$ is the electric charge of electron, $c$ is the speed of light, $m_e$ is the mass of electron, $n_i$ and $n_e$ are ion and electron density. For a $\kappa$ ($\kappa > 3/2$) or Maxwellian ($\kappa \rightarrow \infty$) distribution of electrons, we can substitute the function of the electron distribution and get the free-free emissivity
\begin{equation}
\label{eq:brem}
\begin{split}
&\varepsilon_{ff}(\nu, T) = C_{ff} Z^2 (\frac{T}{1 K})^{-1/2} \\
&  \times \left\{
\begin{array}{r@{\;\;}l}
g_{ff}(\nu, T) e^{-\frac{h \nu}{k_{\rm B}T}},  (\kappa \rightarrow \infty)\\
A_{\kappa}\int^{\infty}_{0} (1+\frac{y+h \nu/k_{\rm B}T}{\kappa-3/2})^{-\kappa-1}g_{ff}(\nu, T)dy,  (\kappa > 3/2)
\end{array}
\right.
\end{split}
\end{equation}
where $C_{ff}=\frac{32\pi^{3/2}e^6}{(3m_e)^{3/2}c^3}\sqrt{\frac{2}{k_B}}n_{i} n_{e}=6.84 \times 10^{-38}n_{i} n_{e}$ erg cm$^{-3}$ s$^{-1}$ Hz$^{-1}$ and the parameter $y$ has the form $y=\frac{E_f-h\nu}{k_BT}$.

\subsection{Radiative Recombination Continuum (RRC)}
Radiative recombination emission, i.e. free-bound radiation, is produced when an electron collides and recombines with an ion, emitting a photon. The photon energy $h\nu$ equals to the kinetic energy of the electron $E_e$ plus the binding energy $I_{Z,j+i}$ of ion at ionization state $j+1$ with the newly-recombined electron, i.e.£¬ $h\nu=E_e+I_{Z,j}$. A sharp edge at the ionization threshold $I_{Z,j}$ of the level will appear in RRC spectrum where the velocity of incident electron is zero. The RRC spectrum of a plasma with many different elements and ions shows many jumps due to recombination into the ground and excited levels of many ions.

For an electron distribution $f(\upsilon, T)$, the RR spectrum due to a photon capturing to an energy level $j$ is (Tucker \& Gould 1966),
 \begin{equation}
 \label{eq:frr0}
 \varepsilon_{fb}(\nu, T)= n_en_{Z,j+1} h\nu \sigma^{rr}_n(\upsilon)f(\upsilon, T) d\upsilon/d(h\nu),
 \end{equation}
where $n_{Z,j+1}$ is the density of ion at the ionization state $j+1$, $\sigma^{rr}_n(\upsilon)$ is the recombination cross section to the considered energy level $n$ at the electron velocity $\upsilon$. It is calculated using the photo-ionization cross section $\sigma^{ph}_n(\nu)=(\frac{m_ec\upsilon}{h\nu})^2\frac{g_{Z,j+1}}{g_{Z,j,n}}\sigma^{rr}_n(\upsilon)$ (Raymond \& Smith 1977) since photo-ionization and recombination are inverse processes and can be related with detailed balance, where $g_{Z,j+1}$ and $g_{Z,j,n}$ are the statistical weights for the ion $(Z, j+1)$ in its ground state and the ion ($Z,j$) in state $n$. After changing the electron distribution $f(\upsilon, T)$ in Eq. (\ref{eq:frr0}) to a function of $\kappa$ or Maxwellian distribution, we can obtain the free-bound emissivity
\begin{equation}
\label{eq:rr}
\begin{split}
&\varepsilon_{fb}(\nu, T) = C_{fb} (\frac{h\nu}{\rm keV})^3(\frac{T}{\rm 1K})^{-3/2} \sigma_{n}^{ph}(h\nu)\\
& \times \left\{
\begin{array}{r@{\;\;}l}
  e^{-\frac{h\nu-I_{Z,j,n}}{k_{\rm B}T}},  (\kappa \rightarrow \infty)\\
 A_{\kappa}(1+\frac{1}{\kappa-3/2}\frac{h\nu-I_{Z,j,n}}{k_{\rm B}T})^{-\kappa-1}, (\kappa > 3/2)
\end{array}
\right.
\end{split}
\end{equation}
where $C_{fb} = 1.31 \times 10^{8} n_{\rm Z,j+1} n_{\rm
  e}\frac{g_{Z,j,n}}{g_{Z,j+1}} $ and $f_{fb}(\nu, T)$ is in the unit of erg cm s$^{-1}$ keV$^{-1}$.
In order to get the RRC emissivity $\varepsilon_{fb}(\nu, T)$ as the function of temperature, we
integrate the above equation (Eq. \ref{eq:rr}) for the photon
energy $h\nu$. Using oxygen ions as an example, we find that in database
AtomDB, we only have photon-ionization cross section $\sigma_{n}^{ph}$
up to n=5 levels for each ion, which is inadequate to capture the entire rate. Therefore, for n$>$5 levels, we approximate the high-n photo-ionization cross sections using the level-dependent RR rates that are in the AtomDB. We calculate the cross sections for those energy levels by assuming the total missing cross section for each ion of oxygen as a single simple edge plus a power-law function. We fit the edge energy, the amplitude and the power law index to match the missing recombination rates. The final RR rate thus can be obtained by integrating the fitted cross section multiplying with Maxwellian or $\kappa$ electron distributions and add the fitted rates to the calculated ones.
\subsection{Dielectronic Recombination (DR)}
Dielectronic recombination occurs where an electron is captured by an ion and simultaneously excites the core. The doubly excited state is unstable and the ion may either auto-ionize (in which case no DR occurs) or radiatively stabilize, emitting a satellite line. The DR process happens only if the kinetic energy of the recombining electron equals to the sum of the energies of the two excited levels $E_c$. DR satellite lines appear to the long-wavelength side of the resonance lines and can also be produced by direct excitation of an inner-shell electron for heavier ions or in transient ionizing plasmas.

The total DR rate coefficients for thermal plasmas are given after averaging over the Maxwellian distribution of the electron energy (e.g., Bates \& Dalgarno 1962; Dubau \& Volonte 1980). Replacing the Maxwellian electron distribution with $\kappa$ distribution shown in Eq. (\ref{eq:Fk}), we can obtain the DR rate coefficients in units of cm$^3$ s$^{-1}$ as follows,
\begin{equation}
\label{eq:dr_rate}
\begin{split}
\alpha_{\rm DR} = &C_{\rm cDR}T^{-3/2}\\
& \times \left\{
\begin{array}{r@{\;\;}l}
e^{-\frac{E_{\rm c}}{k_{\rm B}T}},   (\kappa \rightarrow \infty) \\
A_{\kappa}(1+\frac{E_{\rm c}}{(\kappa-3/2)k_{\rm B}T})^{-\kappa-1}, (\kappa > 3/2)
\end{array}
\right.
\end{split}
\end{equation}
where $C_{\rm cDR}=(\frac{4 \pi R_y}{k_{\rm B}})^{3/2}a_0^{3}V_{\rm a}$, $a_0$ is the Bohr radius in cm, and $V_{\rm a}$ is the capture probability in s$^{-1}$.
Doing the same with the calculations of Kato et al. (1997) for the satellite line intensity in a Maxwellian distribution, the DR line intensity for $\kappa$ and Maxwellian distributions is,
\begin{equation}
\label{eq:dr_density}
\begin{split}
I_{\rm S}(T,l \rightarrow d)& = C_{\rm DR}T^{-3/2}\\
& \times \left\{
\begin{array}{r@{\;\;}l}
e^{-\frac{E_{\rm c}}{k_{\rm B}T}},   (\kappa \rightarrow \infty) \\
A_{\kappa}(1+\frac{E_{\rm c}}{(\kappa-3/2)k_{\rm B}T})^{-\kappa-1}, (\kappa > 3/2)
\end{array}
\right.
\end{split}
\end{equation}
where $C_{\rm DR}=6.60 \times 10^{-24}$ ph cm$^{3}$  s$^{-1}
(\frac{R_y}{k_{\rm B}})^{3/2}V_a n_l n_e$ and $n_l$ the parent ion intensity.

Very similar to the RR rate $\alpha_{RR}$, there are two methods to get the DR rate $\alpha_{DR}$ based on Maxwellian electron energy distribution
from AtomDB. One is the direct extraction from the database and the other is calculated from the Eq. (\ref{eq:dr_rate}) with the exciting energy $E_c$ and capture probability $V_a$ for each ion at different temperatures from AtomDB. At the same time, the rate from $\kappa$ electron distribution is corresponding to two methods. One is from the decomposed Maxwellian distribution and the other is the calculation with Eq. (\ref{eq:dr_rate}). Unlike with the RR case, there is no obvious simple top up function that can be easily applied to the direct calculation to account for the incomplete DR satellite line data held in AtomDB, so we expect the DR rates to be under-counted in the direct calculation.

\subsection{Charge State Distribution (CSD)}
 In this work, we assume the CSD is not evolving in time, i.e. we are in collisional ionization equilibrium (CIE). In a CIE plasma, the relative ion abundance is determined by the balance between the electron impact ionization and electron-ion recombination rates out of each charge state. To examine the effect of Kappa electron energy distributions on the CSD, we extract the Maxwellian ionization, $S_{ci}$, and  recombination, $\alpha_{DR}, \alpha_{RR}$, rate coefficients of oxygen and then use the Maxwell decomposition method to calculate $\kappa$ rate coefficients, $S_{ci}^{\kappa}, \alpha_{RR}^{\kappa}, \alpha_{DR}^{\kappa}$. We compare to the results from directly calculated recombination rate coefficients and to those given by KAPPA package.

\begin{figure*}
\centering
  \includegraphics[width=0.9\textwidth]{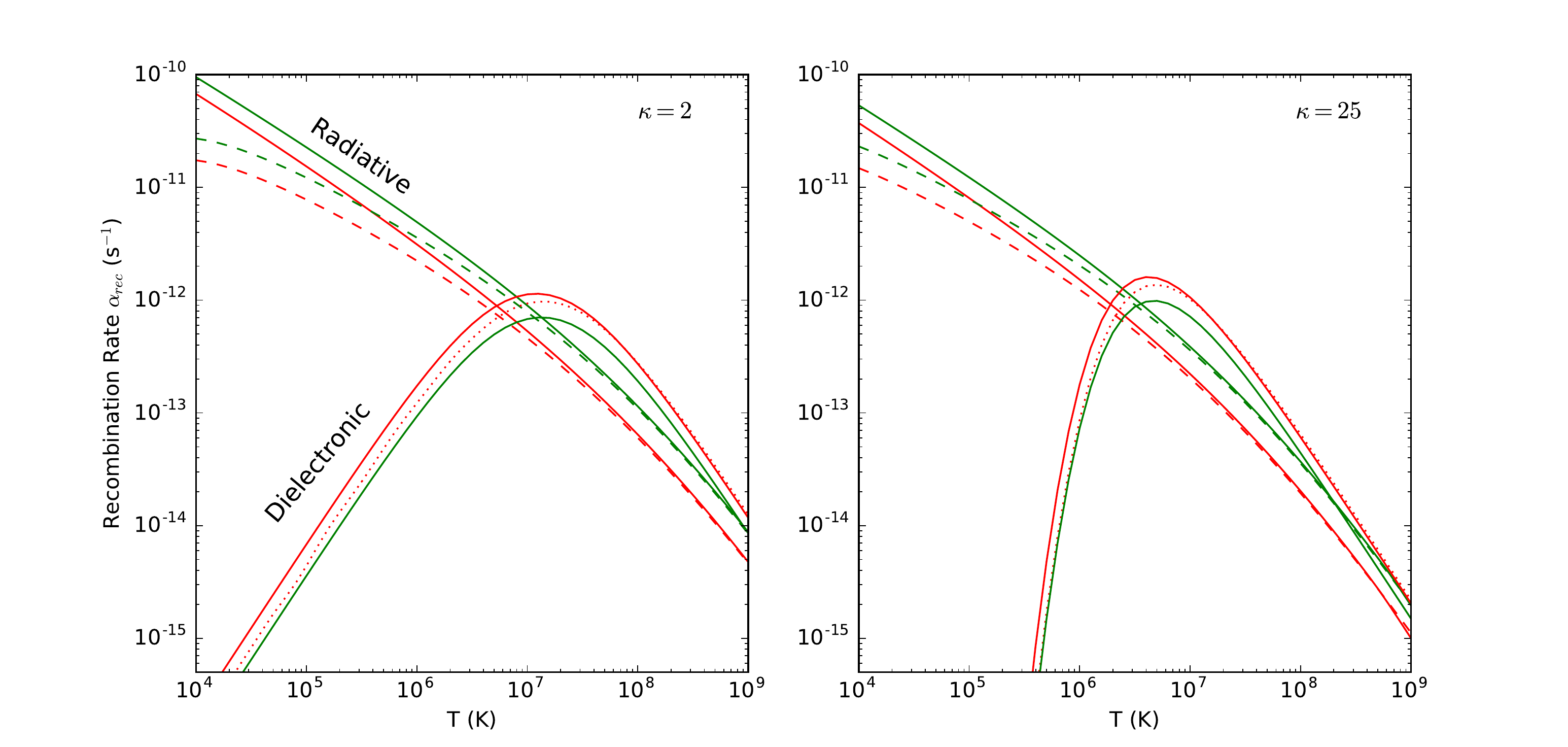}
  \caption{Recombination rates $\alpha_{rec}$ (including RR and DR rates) of oxygen ions versus the temperature for the $\kappa$ distributions with $\kappa=2$ (left panel) and $\kappa=25$ (right panel). The red lines are the rate coefficients of O VII and green lines are those of O VIII. The solid lines are the RR rates and DR rates decomposed from AtomDB. The dotted lines and dashed ones are the analytical rate coefficients.}
  \label{fig:recomb_rate}
\end{figure*}

We plot the recombination rate coefficients as the function of temperature for the $\kappa$ distributions in Fig. \ref{fig:recomb_rate}. Left and right panels correspond to $\kappa=2$ and $\kappa=25$ cases. The red and green lines are the rate coefficients for O VII and O VIII, respectively. The solid lines are RR rates and DR rates decomposed from the Maxwellian RR rates in AtomDB. The dotted and dashed lines are RR and DR rates from analytical calculations. The analytical DR rate is lower than the decomposition due to the incomplete DR set of DR satellite lines in AtomDB, discussed above.

\begin{figure*}
\centering
\includegraphics[width=0.9\textwidth]{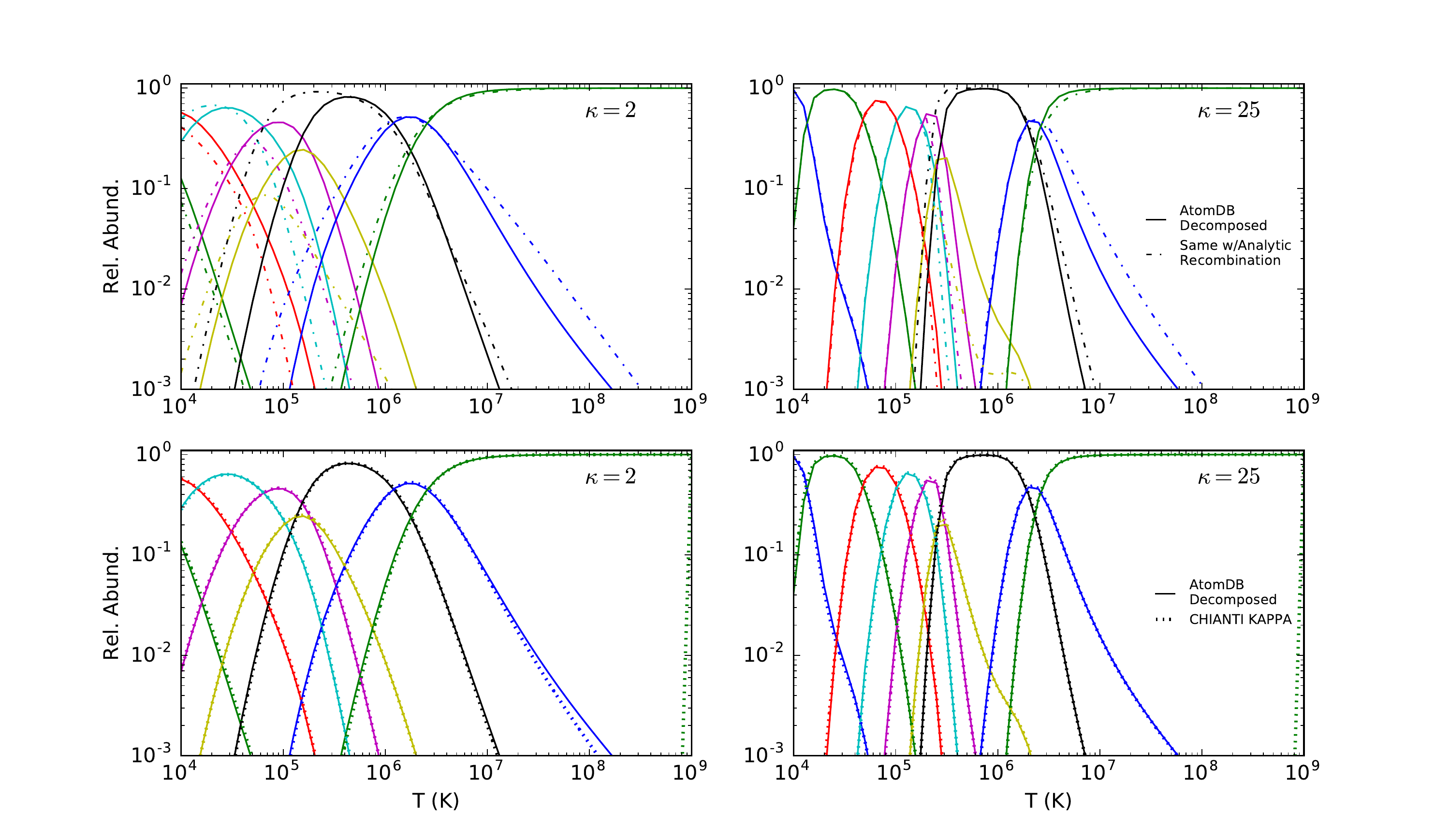}
  \caption{The CSD for oxygen as a function of temperature for $\kappa$ distributions with $\kappa=2$ (left panels) and $\kappa$=25 (right panels). The solid lines are the oxygen CSD from rate coefficients decomposed from the data in AtomDB. The dash dot
  lines are according to the relative abundance from the analytical recombination rates. For the rates we can not obtain from analytical method, we still use the coefficients from the Maxwellian decomposition method. The dotted lines are the results from the KAPPA database with available ion data in CHIANTI database.}
  \label{fig:csd_decomp_cal}
\end{figure*}

We then used the Maxwell decomposition method on all ions, and determine the CSD (solid lines in Fig. \ref{fig:csd_decomp_cal}). The dash dot lines in this figure are the results from the analytical calculations of recombination rates. The analytical method can give the rate of O VII that can not be obtained from the decomposed method. For the rates that we can not obtain from analytical method (there is no DR satellite line information in AtomDB for O I-VI), we use the decomposition methods. We plot the result from the KAPPA package with dotted lines for the comparison in the lower panels of Fig. \ref{fig:csd_decomp_cal}. The analytic results differ from the decomposition and KAPPA package results due to the inadequacies of the RR approximation. But the decomposed results well consistent with those from the KAPPA package.
\section{Method} \label{Spectral analysis}
Calculating the spectrum of a non-Maxwellian plasma requires several types of non-Maxwellian data. As described in section \ref{proce}, several continuum components can be modeled from first principles as non-Maxwellians. However, the majority of relevant atomic data is published on a Maxwell-averaged temperature grid, as this both saves space requirements and is usually the relevant electron distribution. This is especially true for line emission: the collisional excitation cross section has many resonances in the cross sections, and accurately modeling a new electron distribution would require re-integrating the original energy dependent cross sections with the new distribution; in most cases these energy resolved files no longer exist and their regeneration would be impractical. Therefore, we adopt the decomposition approach of Hahn \& Savin (2015) to create spectra using AtomDB, by adding Maxwellian spectra.

The AtomDB non-equilibrium ionization data provides the line and continuum emissivity for each ion as a function of photon energy (0.01 to 100.0 keV) and plasma temperature ($10^4$ to $10^9$K). The elemental abundances from Andres \& Grevesse (1989) are built in to the emissivities, but the ion fractions are not. To create a Maxwellian spectrum, these emissivities must be multiplied by the ion fraction. By default the ionization and recombination rates of Bryans et al. (2009) are used in AtomDB.

To create the $\kappa$ emissivity, firstly the ionization fraction is calculated using $\kappa$ ionization and recombination rates (obtained by decomposition). Then the emissivities are calculated for each ion (also calculated by decomposition) and the emissivity is multiplied by the relevant ion fraction to get the total emissivity.

Nozawa et al. (2009) gave accurate analytic fitting formulae for the e-e bremsstrahlung in a wider electron temperature range 1keV $\leqslant$ $k_B$T $\leqslant$ 1MeV. We apply their results and add e-e thermal bremsstrahlung calculations to APEC model.

Maxwellian decomposition normalization parameters ($a_j$ in Table 7 of paper Hahn \& Savin 2015) require Maxwellian temperatures out of the range of [10$^4$, 10$^9$]K. We define a ``bad'' number ratio of $a_j$ in those regions to the total number, $\frac{\sum a_{j,bad}}{\sum a_{j}}$,  in the space
of $\kappa$ and temperature $T$, which implies the probability of this method for the conversion between two distributions being a bad approximation. Here, $\sum a_{j,bad}$ is the number of the $a_j$ out side of the temperature range [10$^4$ 10$^9$]K. The larger ratio shows the more invalid of Maxwellian decomposition method.

\begin{figure*}
\centering
\includegraphics[width=0.5\textwidth]{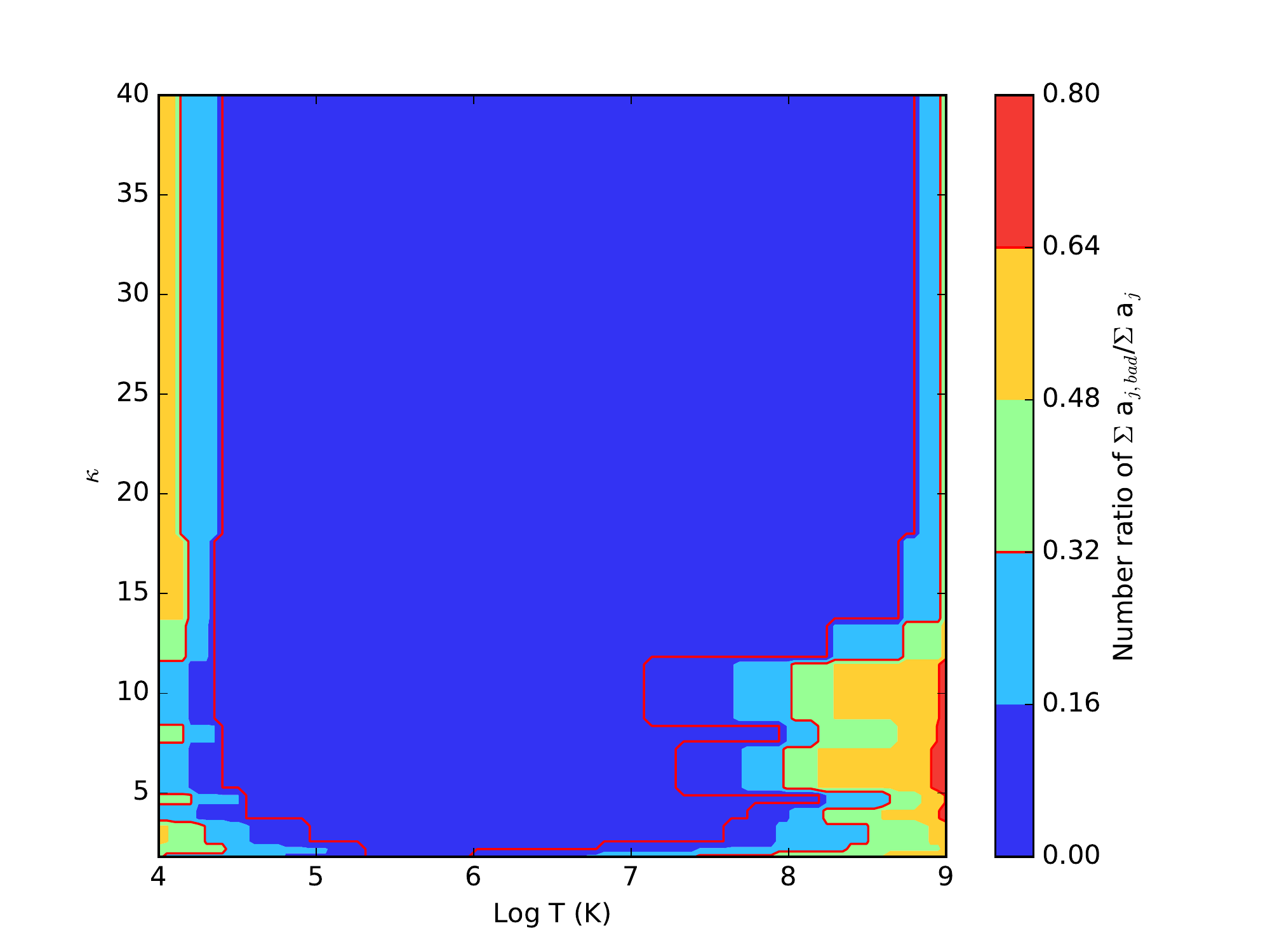}
  \caption{The ``bad'' number ratio $\frac{\sum a_{j,bad}}{\sum
      a_{j}}$ of the converting parameter $a_j$ for Maxwellian
    decomposition method. The ``bad'' number $\sum a_{j,bad}$ shown in color bar is defined as the number of Maxwellian temperature outside of the range of [$10^4$ $10^9$] K when decomposing $\kappa$ temperature to a temperature matrix of Maxwellian before sum the weighted Maxwellian rate to get the $\kappa$ rate. The larger ratio corresponds to the more invalid of Maxwellian
decomposition method. It shows large ratios when temperatures approach to the two boundaries ($10^4$ and $10^9$ K) and electrons distributions toward to the non-thermal.}
\label{fig:contourg}
\end{figure*}

We plot the ``bad'' number ratio $\frac{\sum a_{j,bad}}{\sum a_{j}}$ as the function of index $\kappa$ and temperature $T$ in Fig. \ref{fig:contourg}. From the figure, we can find the probability and limits of Maxwellian decomposition for the $\kappa$ rate coefficient calculations. When the logarithm of temperature are less than 4.5 and larger than about 8.5, the invalid probability of the decomposition method is higher as the temperature tends to the boundaries. The ``bad'' number ratio is larger at higher temperature if $\kappa$ is less than about 15.

\section{Results} \label{Results and Discussion}
\subsection{Spectra from $\kappa$ Distributions}
Applying the Maxwellian decomposition method described in Sect \ref{proce}, we produce bremsstrahlung emissivity from $\kappa$-distributed electrons and compare with those from analytical calculation. We use the deconvolution method to get the $\kappa$ CSDs for oxygen from AtomDB data.

\begin{figure*}
\centering
\includegraphics[width=0.6\textwidth]{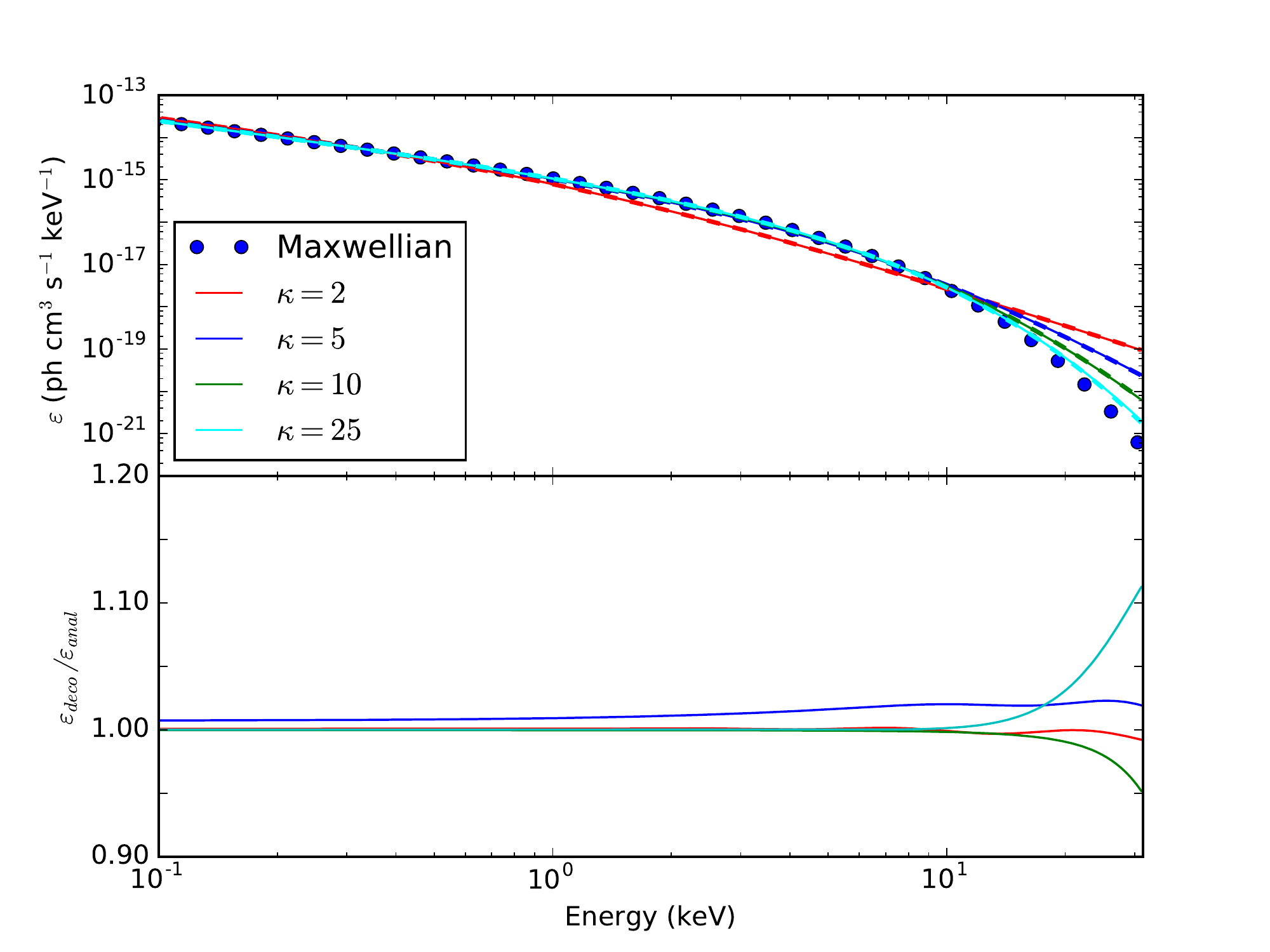}
\caption{The bremsstrahlung emissivity at temperature 3 keV. Upper panel, the free-free continuum emission from $\kappa$ distribution of electrons (lines) and that from Maxwellian distribution (solid blue circles). Dashed lines are from the analytical calculations with Eq. \ref{eq:brem} and solid lines are from Maxwellian decomposition method, respectively. The color of the line shows the result from different $\kappa$ distribution. Lower panel, the emissivity ratio between that from decomposition method $\varepsilon_{\rm deco}$ and that from analytical calculations $\varepsilon_{\rm anal}$. The colors of the lines presents the $\kappa$ values shown in upper panel. }
\label{fig:bremAna}
\end{figure*}

\begin{figure*}
\centering
\includegraphics[width=0.6\textwidth]{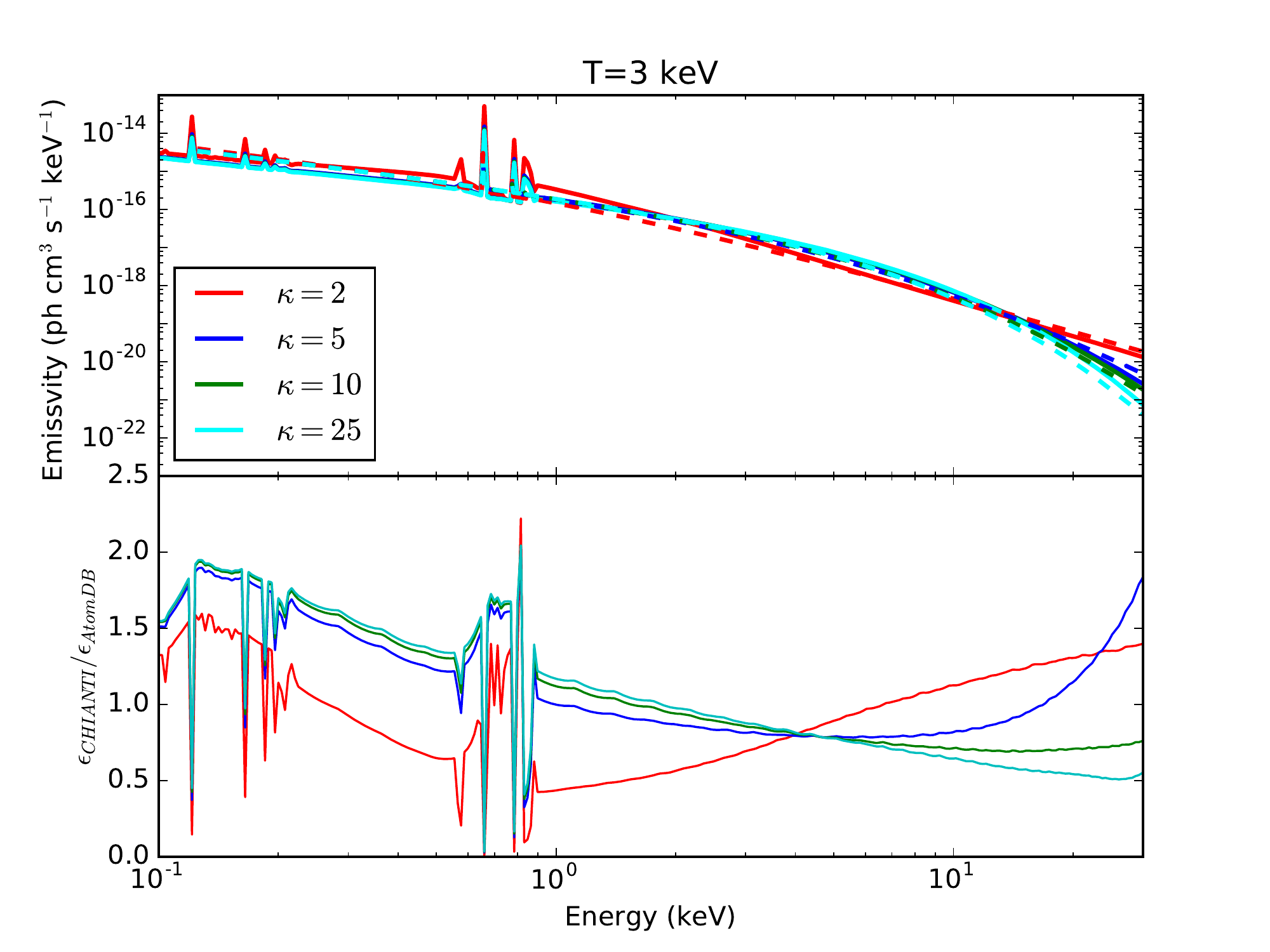}
  \caption{Upper panel: The oxygen emissivity at temperature 3 keV from CHIANTI KAPPA package (dashed lines) and those from AtomDB (solid lines) for $\kappa=$2, 5, 10, 25. Lower panel: The emissivity ratio $\varepsilon_{CHIANTI}/\varepsilon_{AtomDB}$ between that from KAPPA package (dashed lines in upper panel) with CHIANTI data and that from AtomDB data (solid lines in upper panel). The dash line shows the ratio $\varepsilon_{CHIANTI}/\varepsilon_{AtomDB}=1$. The color of solid line corresponds to $\kappa$ value as that in upper panel.}
\label{fig:oxy_Chianti}
\end{figure*}

Thermal bremsstrahlung produced by electron-ion interactions are the dominant source of energy loss in an optically thin hot plasma. Using Eq. (\ref{eq:brem}), we obtain the free-free emissions with Maxwellian and $\kappa$ distributions of electrons at temperature of 3 keV. There are three methods used in the APEC codes for bremsstrahlung continuum calculations, including the relativistic case (Nozawa et al. 1998), the non-relativistic case (Hummer 1988), and a semi-relativistic case (Kellogg, Baldwin \& Koch 1975). With the relativistic approach, we plot the analytical bremsstrahlung emission with dashed lines and decomposed ones with solid lines in Fig. \ref{fig:bremAna}. The emissivity from a Maxwellian electron distribution at 3 keV is plotted with solid circles. We find that both methods are almost consistent for $10^{-2}$keV$< E < 30$keV.
In order to obtain $\kappa$-distribution spectra from Maxwellian decomposition method, we take oxygen as an example and extract the spectrum of each oxygen ion from AtomDB. We combine line emission and continuum for each ion as the total emission of this ion. For the $\kappa$-distributed electron and given temperature $T_\kappa$, we firstly decompose the ion spectrum extracted from AtomDB NEI data (line and continuum) based on $\kappa$ value and Maxwellian decomposition method. Secondly, we select the ion abundance for each ion at the nearest temperature with $T_\kappa$. Thirdly, we multiply the decomposed emissivity with selected ion abundance to get the NEI emissivity for each ion. The $\kappa$ emissivity for oxygen finally obtained with the sum of emission from each ion of oxygen. We show the total spectra (solid lines) from $\kappa$ distributions with $\kappa$=2, 5, 10, and 25 in Fig. \ref{fig:oxy_Chianti}. We also plot the oxygen spectra of $\kappa$ distributions from the KAPPA package as dashed lines in this figure.
We find the line emissivities from $\kappa=2$ are larger than those from other $\kappa$ distributions. The continuum emission from $\kappa=2$ is higher at lower ($<$2 keV) and higher ($>$20 keV) energies.

\subsection{Application of e-e Bremsstrahlung}
\begin{figure*}
\centering
\includegraphics[width=0.6\textwidth]{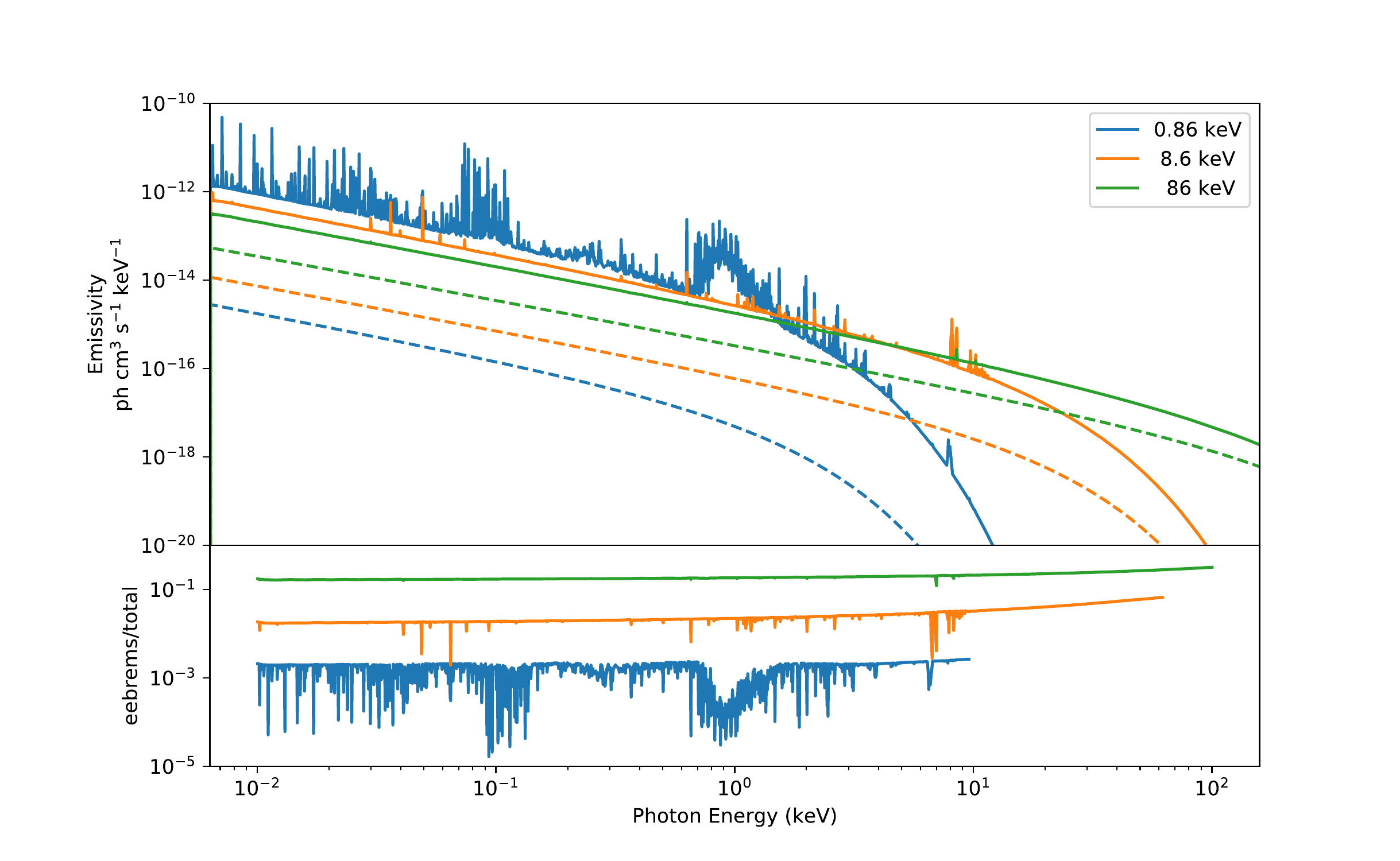}
  \caption{Upper panel, the total emissivity
    (solid lines) and the e-e thermal bremsstrahlung emissivity (dot
    lines) at three temperatures, $k_B$T=0.86keV (blue lines), 8.6keV (orange
    lines), and 86keV (green lines). E-E thermal bremsstrahlung emissivity is from the
    work of Nozawa et al. (2009). Lower panel, the emissivity ratio
    between e-e thermal bremsstrahlung and total spectra at three
    temperatures in upper panels.}
\label{fig:eebrem}
\end{figure*}

With the updated APEC model including e-e bremsstrahlung emission, we plot the total emissivity, the e-e thermal bremsstrahlung emissivity, and their ratio at three temperatures ($k_B$T=0.86keV, 8.6keV, and 86keV) in Fig. \ref{fig:eebrem}. We find the e-e contribution becomes larger at higher temperature. At $k_B$T=86keV (i.e., 10$^9$K, green line in lower panel), e-e bremsstrahlung emission is $>$10\% of the total emissions. But at $k_B$T=0.86keV (blue line in lower panel), the contribution is less than 0.1\% at energy less than about 1keV.
As a practical example of high temperature plasma where the inclusion of e-e bremsstrahlung has non-negligible effect, we consider hot plasma contained in a PSR of a magnetic cataclysmic variable. Among them, intermediate polar (IP) type CV which is thought to have magnetic field of order B $\approx 10^5-10^7$ G (de Martino et al. 2004) generally shows harder X-ray spectra than those of polars (B $\approx 10^7-10^9$ G) due to less cyclotron cooling suggesting higher plasma temperature in the PSR. To estimate the mass of a white dwarf (WD) in IPs based on the spectral fitting, several authors developed a numerical model of the PSR and applied it to spectra obtained with multiple X-ray telescopes (Cropper et al. 1999, Suleimanov et al. 2005, Yuasa et al. 2010). In these models, accreting gas is assumed to freely fall onto a WD along the magnetic field line inside the magnetosphere, and at a certain height (the shock height), the bulk kinetic energy is converted to thermal energy, and the gas is heated to high temperatures (typically $k_{\rm B}$T$>$a few $\times$ 10 keV). The shock-heated gas is cooled via optically-thin thermal emission, and lands on to the atmosphere of the WD. The previous model numerically solved hydrodynamic equations along the PSR, and generated total X-ray spectra based on various plasma models; e.g. bremsstrahlung and empirical gaussian were used by Suleimanov et al. (2005), Yuasa et al. (2010) adopted APEC to more accurately treat He-like and H-like Fe K emission lines.

\begin{figure*}
\centering
\includegraphics[width=0.45\textwidth]{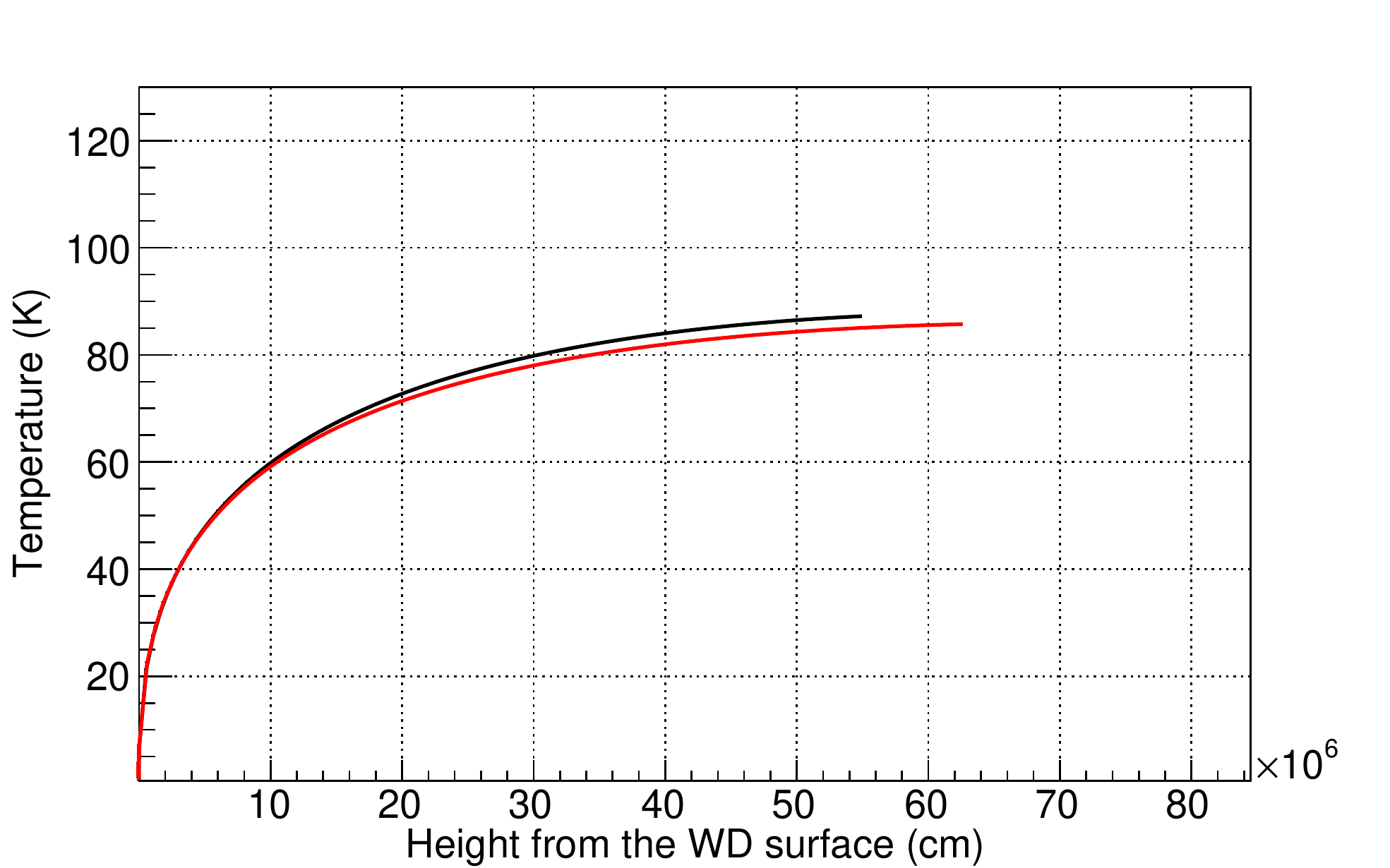}
\includegraphics[width=0.45\textwidth]{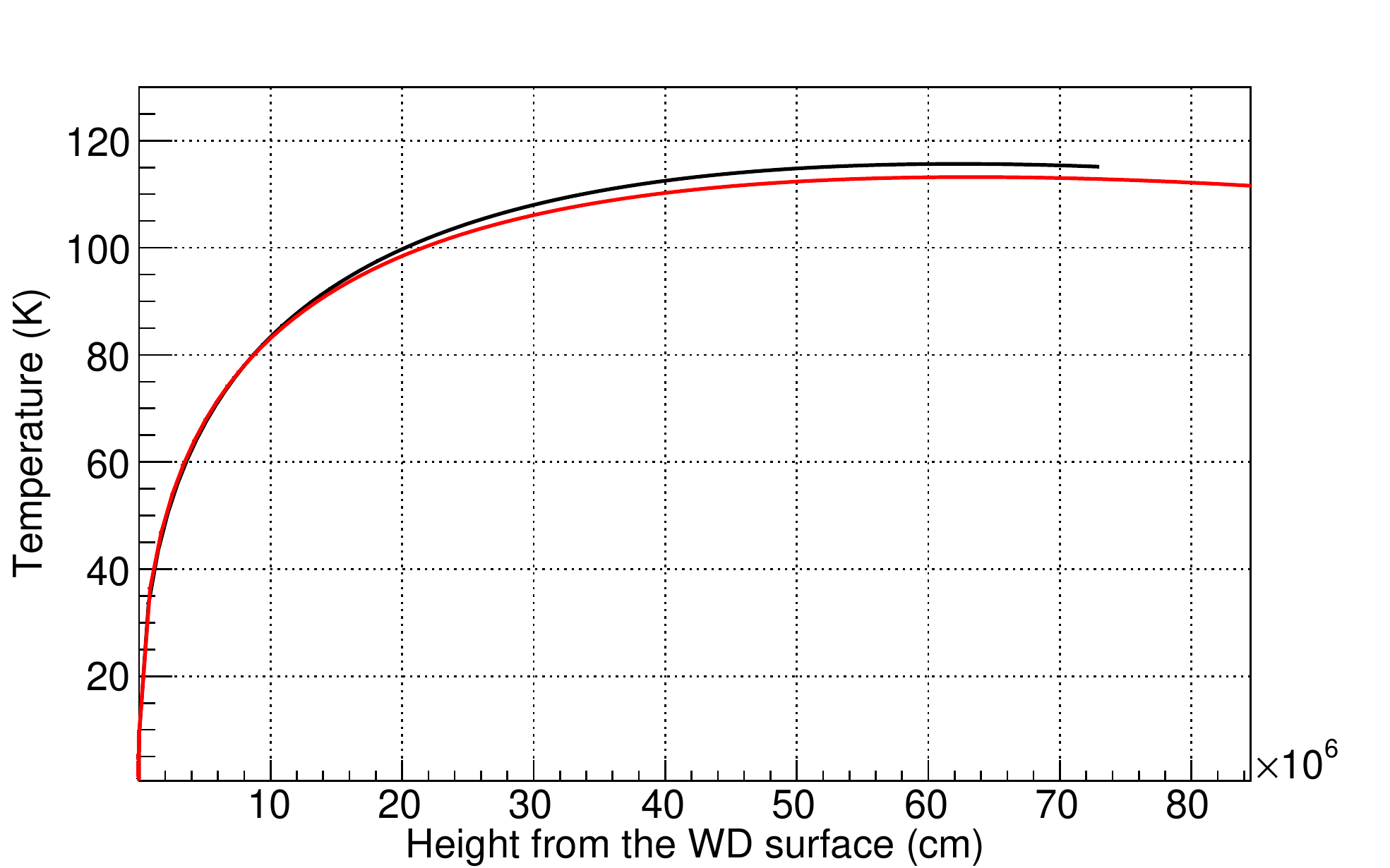}
\includegraphics[width=0.45\textwidth]{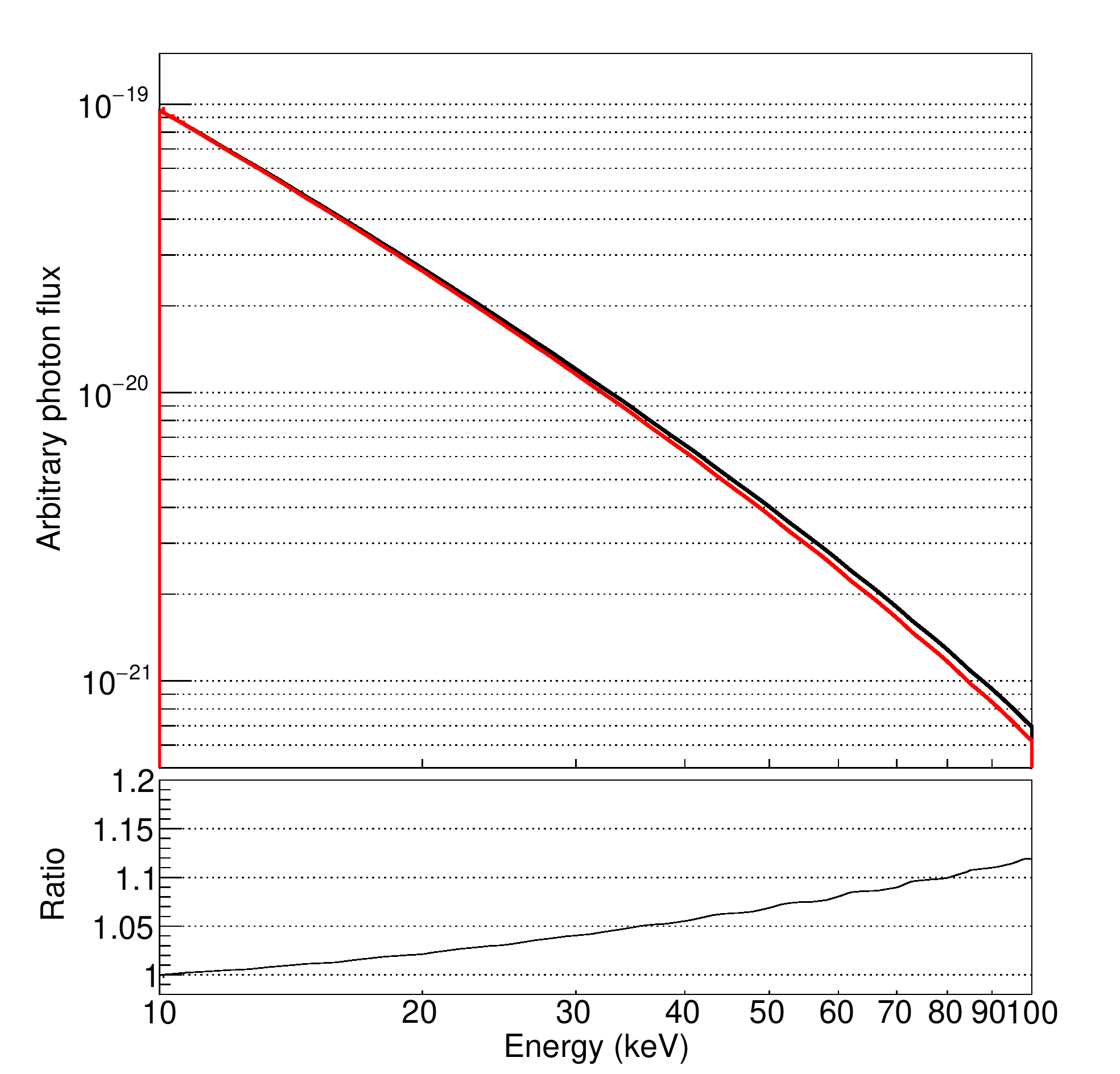}
\includegraphics[width=0.45\textwidth]{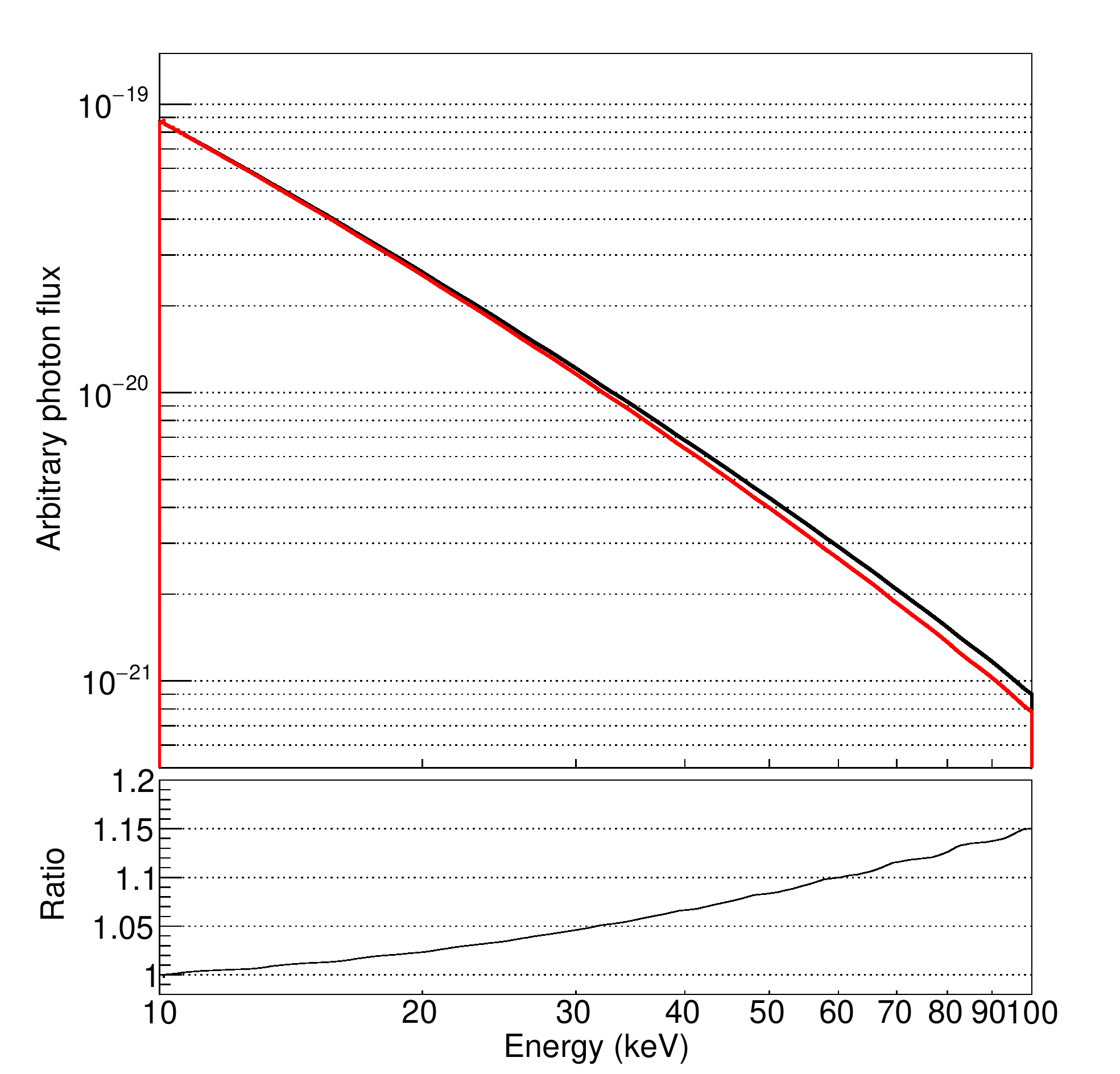}
  \caption{Plasma temperature profiles (upper panels) and total spectra (lower panels) from the PSR calculated from the updated numerical hydrostatic model for an IP of mass $M_{\rm WD}$=1.2 $M_\odot$ (left panel) and $M_{\rm WD}$=1.3 $M_\odot$ (right panel). Red and black solid lines are corresponding to the updated APEC model with and without e-e thermal bremsstrahlung.}
\label{fig:WDfit}
\end{figure*}

Applying the updated APEC model including e-e thermal bremsstrahlung emissions to the numerical PSR model of Yuasa et al. (2010), we calculate the temperature gradient in the PSR of an IP with a WD mass of MWD=1.2 M$_{\odot}$ and that with M$_{\rm WD}$=1.3 M$_{\odot}$ under the assumptions and the boundary conditions described therein. Table 1 summarizes shock height and shock temperature obtained with and without considering e-e bremsstrahlung. The upper panels in Fig. \ref{fig:WDfit} show the calculated temperature profile along the PSR height. When e-e bremsstrahlung is included, plasma more effectively cools in higher temperatures  ($\gtrsim 10^8$~K), and therefore a higher shock temperature can satisfy the boundary conditions which requires $k_{\rm B}$T=0 keV at the WD surface. This leads to a lower shock height and higher shock temperature with e-e bremsstrahlung than without. The shock temperatures differ from each other by 2\% (M$_{\rm WD}$=1.2 M$_{\odot}$) and 3\% (M$_{\rm WD}$=1.3 M$_{\odot}$), and thus, the inclusion of e-e bremsstrahlung does not significantly change the vertical structure of the PSR plasma, nor the emissivity gradient. Rather, because the relative contribution from e-e bremsstrahlung depends on energy, and is larger at higher energies (Fig. \ref{fig:eebrem}) , the change in the continuum spectrum has a larger impact when composing a total PSR emission spectrum  and comparing it with observed data. As shown in the lower panels of Fig. \ref{fig:WDfit}, the total emissivities for M$_{\rm WD}$=1.2 M$_{\odot}$ and 1.3 M$_{\odot}$ differ by $>10\%$ in energies $E>75$~keV and $E>60$~keV, respectively, when two spectra with and without e-e bremsstrahlung are normalized at $E=10$~keV.
\begin{table}
\caption{The numerical solutions for the shock height $h_{\rm PSR}$ of PSR and the shock temperature $T_h$ at height $h_{\rm PSR}$.}
\label{tab:table1}
  \begin{tabular}{|l|l|l|l|l|l|l|}
    \hline
    $M_{\rm WD}$ &
      \multicolumn{2}{c|}{1.2$M_{\odot}$} &
      \multicolumn{2}{c|}{1.3$M_{\odot}$} \\
      \hline
    APEC& with e-e & without e-e  & with e-e  & without e-e \\
    \hline
    $h_{\rm PSR}$ (km) & 550 &626 & 730 & 845 \\
    \hline
    $T_h$ (keV) & 87.2 & 85.7 & 115 & 112 \\
    \hline
  \end{tabular}
\end{table}

\section{Conclusion}
With released database AtomDB and python module ``Pyatomdb'', we study the X-ray spectra with a non-thermal $\kappa$-distribution of electrons and e-e bremsstrahlung in collisionally ionized plasma. A Maxwellian Decomposition is applied to calculate the coefficient rates of $\kappa$ distributions from the appropriately weighted Maxwellian rates stored in our AtomDB database. We added the e-e bremsstrahlung process into APEC model since a very hot plasma can also produce energetic electrons. We use the updated APEC model to calculate the X-ray spectra from a numerical PSR model of a magnetic cataclysmic variable and find more than 10\% difference for total emissivity with and without e-e bremsstrahlung. We compare the oxygen charge state distribution and spectra to the results from the KAPPA package and find the caution has to be taken when decomposing Maxwellian rate for $\kappa$-distributed coefficients when plasma temperature approached to the boundaries of [10$^4$ 10$^9$] K and the $\kappa$ less than about 15. We will be creating an XSPEC model for the spectral analysis with the $\kappa$ distribution.%

\acknowledgments

We are grateful to Maxim Markevitch for the initial impetus to look at this question, and to Michael Hahn and Daniel Savin for association with their decomposition method. XHC acknowledges financial supports from China Scholarship Council, the National Key R\&D Program of China (Grant No: 2018YFA0404202) and the National Natural Science Foundation of China (U1738125 \& U1938112). RKS and AF acknowledge support from Chandra grant \#TM6-17009X for this work.

\clearpage

\begin{thebibliography}{}
\bibitem[Anders \& Grevesse (1989)]{ande89} Anders, E. \& Grevesse, N. 1989, GeCoA, 222, 307
\bibitem[Barbosa et al.(2016)]{barb16} Barbosa, A., Alves, M. V. \& Sim\~{o}es J. F. 2016, EGUGA, 18, 8124
\bibitem[Bates \& Dalgarno (1962)]{bate62} Bates, D. R. \& Dalgarno, A. 1962, in ``Atomic and Molecular Processes'',
ed. D. R. Bates (New York: Academic Press), 258
\bibitem[Bryans et al.(2009)]{brya09} Bryans, P., Landi, E. \& Savin, D. W. 2009, \apj, 691, 1540
\bibitem[Burgess \& Tully (1992)]{burg92} Burgess, A. \& Tully, J. A. 1992, A\&A, 254, 436
\bibitem[Bykov et al. (2013)]{byko13} Bykov, A. M., Malkov, M. A., Raymond, J. C. et al. 2013, SSRv, 178, 599
\bibitem[Carbary et al. (2014)]{carb14} Carbary, J. F., Kane, M., Mauk, B. H. \& Krimigis, S. M. 2014, JGRA, 119, 8426
\bibitem[Christon (1987)]{chis87} Christon, S. P. 1987, Icarus, 71, 448
\bibitem[Collier et al.(1996)]{coll96} Collier, M. R., Hamilton, D. C., Gloeckler, G., Bochsler, P., \& Sheldon, R. B.
1996, GeoRL, 23, 1191
\bibitem[Collier (2004)]{coll04} Collier, M. R. 2004, AdSpR, 33, 2108
\bibitem[Cropper et al.(1999)]{crop99} Cropper, M., Wu, K., Ramsay, G. et al. 1999, \mnras, 306, 684
\bibitem[de Avillez \& Breitschwerdt(2015)]{deav15} de Avillez, M. A. \& Breitschwerdt, D. 2015, A\&A, 580, 124
\bibitem[de Martino et al.(2004)]{dema04} de Martino, D., Matt, G., Belloni, T., et al. 2004, Nucl. Phys. B Proc. Suppl.,
132, 693
\bibitem[De Pontieu et al.(2014)]{depo14} De Pontieu, B., Title, A. M., Lemen, J. R., et al. 2014, SoPh., 289, 2733
\bibitem[Dialynas et al. (2009)]{dial09} Dialynas, K., Krimigis, S. M., Mitchell, D. G., Hamilton, D. C., Krupp, N. \& Brandt P. C. 2009, J. Geophys.
Res., 114, A01212
\bibitem[Dubau \& Volonte(1980)]{duba80} Dubau, J., \& Volonte, S. 1980, Rep. Prog. Phys., 43, 199
\bibitem[Dud\'ik (2015)]{dud15} Dud\'ik, J., Mackovjak, \v{S}, Dzif\v{c}\'akov\'a, E., et al. 2015, \apj, 807, 123
\bibitem[Dud\'ik (2017)]{dud17} Dud\'ik, J., Dzif\v{c}\'akov\'a, E., Meyer-Vernet, N., et al. 2017, SoPh, 292, 100
\bibitem[Dzif\v{c}\'akov\'a(1992)]{dzi92} Dzif\v{c}\'akov\'a, E. 1992, SoPh, 140, 247
\bibitem[Dzif\v{c}\'akov\'a \& Kulinov\'a (2010)]{dzi10} Dzif\v{c}\'akov\'a, E., \& Kulinov\'a, A. 2010, SoPh, 263, 25
\bibitem[Dzif\v{c}\'akov\'a et al.(2011)]{dzi11} Dzif\v{c}\'akov\'a, E., Homola, M., \& Dud\'ik, J., 2011, A\&A, 531, 111
\bibitem[Dzif\v{c}\'akov\'a \& Dud\'ik(2013)]{dzi13} Dzif\v{c}\'akov\'a, E., \&  Dud\'ik, J. 2013, \apjs, 206, 9
\bibitem[Dzif\v{c}\'akov\'a et al.(2015)]{dzi15} Dzif\v{c}\'akov\'a, E., Dud\'ik, J., Kotr\v{c}, P. et al., 2015, \apjs, 217, 14
\bibitem[Dzif\v{c}\'akov\'a et al.(201)]{dzi18} Dzif\v{c}\'akov\'a, E., Zemanov\'a, A., Dud\'ik, J. \& Mackovjak, \v{S}. 2018, \apj, 853, 158
\bibitem[Foster et al.(2012)]{fost12} Foster, A. R., Ji, L., Smith, R. K., \& Brickhouse, N. S. 2012, \apj, 756, 128
\bibitem[Hahn \& Savin(2015)]{hahn15} Hahn, M., \& Savin, D. W. 2015, \apj, 800, 68
\bibitem[Hasegawa et al. (1985)]{Hasegawa85} Hasegawa, A., Mima, K. \& Duong-van, M. 1985, Phys. Rev. Lett., 54, 2608
\bibitem[Hasegawa (1989)]{Hasegawa89} Hasegawa, A., \& Sato, T. 1989, Space Plasma Physics, Vol. 1 (Berlin: Springer)
\bibitem[Haug (1975a)]{haug75a} Haug, E. 1975a, SoPh, 45, 453
\bibitem[Haug (1975b)]{haug75b} Haug, E. 1975b, ZNatA, 30, 1546
\bibitem[Hummer(1988)]{humm88} Hummer, D. G. 1988, \apj, 327, 477
\bibitem[Henning et a. (2011)]{henn11} Henning, F. D., Mace, R. L., \& Pillay, S. R. 2011, JGRA, 11612203
\bibitem[Itoh et al.(1985)]{itoh85} Itoh, N., Nakagawa, M., \& Kohyama, Y. 1985, \apj, 294, 17
\bibitem[Itoh et al.(1990)]{itoh90} Itoh, N. Kojo, K. \& Nakagawa, M. 1990, \apjs, 74, 291
\bibitem[Itoh et al.(2000)]{itoh00} Itoh, N., Sakamoto, T. Kusano, S. et al. 2000, \apjs, 128, 125
\bibitem[Jeffrey (2016)]{jeff16} Jeffrey, N. L. S., Fletcher, L., \& Labrosse, N. 2016, A\&A, 590, A99
\bibitem[Jeffrey (2017)]{jeff17} Jeffrey, N. L. S., Fletcher, L., \& Labrosse, N. 2017, \apj, 836, 35
\bibitem[Kaastra, Bykov, \& Werner(2009)]{kaas09} Kaastra, J. S., Bykov, A. M., \& Werner, N. 2009, A\&A, 503, 373
\bibitem[Karzas\& Latter(1961)]{karz61} Karzas, W. J. \& Latter, R. 1961, \apjs, 6, 167
\bibitem[Ka\v{s}parov\'a\& Karlick\'y (2009)]{kasp09} Ka\v{s}parov\'a, J. \& Karlick\'y, M. 2009A\&A, 497, 13
\bibitem[Kato et al.(1997)]{kato97} Kato, T, Safronova, U. I., Shlyaptseva, A. S., Cornille, M., Dubau, J. \&
Nilsen, J. 1997, ADNDT, 67, 225
\bibitem[Kellogg et al.(1975)]{kell75} Kellogg, E., Baldwin, J. R., \& Koch, D. 1975, \apj, 199, 299
\bibitem[Ko et al.(1996)]{ko96} Ko, Y. K., Fisk, L. A. Gloeckler, G. et al., 1996, Geophys. Res. Lett., 23, 2785
\bibitem[Kowk(2007)]{kwok07} Kwok, S. 2007, Physics and Chemistry of the Interstellar Medium (Sausalito:
University Science Books)
\bibitem[Le Chat et al. (2011)]{lech11} Le Chat, G., Issautier, K., Meyer-Vernet, N., \& Hoang, S. 2011, SoPh,
271, 141
\bibitem[Leubner (2002)]{leub02} Leubner, M. P. 2002, Ap\&SS, 282, 573
\bibitem[Lin et al. (2002)]{lin02} Lin, R. P., Dennis, B. R., Hurford, G. J., et al. 2002, SoPh., 210, 3
\bibitem[Livadiotis(2018)]{liva18} Livadiotis, G. 2018, Univ, 4, 144
\bibitem[Livadiotis \& McComas(2009)]{liva09} Livadiotis, G., \& McComas, D. J. 2009, J. Geophys. Res., 114, A11105
\bibitem[Livadiotis \& McComas(2010)]{liva10} Livadiotis, G., \& McComas, D. J. 2010, \apj, 714, 971
\bibitem[Livadiotis \& McComas(2013)]{liva13} Livadiotis, G., \& McComas, D. J. 2013, SSRv, 175, 215
\bibitem[Livadiotis (2018)]{liva18}Livadiotis, G., Desai, M. I., \& Wilson, L. B., III 2018, \apj, 853, 142
\bibitem[Mace \& Hellberg(2013)]{mace95} Mace, R. L. \& Hellberg, M. A. 1995, PhPl, 2, 2098
\bibitem[Mackovjak et al. (2013)]{mack04} Mackovjak, \v{S}, Dzif\v{c}\'akov\'a, E., \& Dud\'ik, J. 2013, SoPh, 282, 263
\bibitem[Mauk et al. (2004)]{mauk04} Mauk, B. H., et al. 2004, J. Geophys. Res., 109, A09S12
\bibitem[Maxon \& Corman (1967)]{maxo67} Maxon, M. S. \& Corman, E. G. 1967, Phys. Rev. 163, 156
\bibitem[Meyer-Vernet et al. (1995)]{meye95} Meyer-Vernet, N., Moncuquet, M., \& Hoang, S. 1995, Icar, 116, 202
\bibitem[Nicholls et al. (2012)]{nich12} Nicholls, D. C., Dopita, M. A., \& Sutherland, R. S. 2012, \apj, 752, 148
\bibitem[Nozawa Itoh \& Kohyama(1998)]{noza98} Nozawa, S., Itoh, N. \& Kohyama, Y. 1998, \apj, 507, 530
\bibitem[Nozawa et al. (2009)]{noza09} Nozawa, S., Takahashi, K., Kohyama, Y., \& Itoh, N. 2009, A\&A, 499, 661
\bibitem[Olbert et al.(1967)]{olbe67} Olbert, S., Egidi, A., Moreno, G., \& Pai, L. G. 1967, Trans. AGU, 48, 177
\bibitem[Owocki \& Scudder(1983)]{owok83} Owocki, S. P., \& Scudder, J. D. 1983, \apj, 270, 758
\bibitem[Petrosian \& East (2008)]{petr08} Petrosian, V. \& East, W. E. 2008, \apj, 682, 175
\bibitem[Phillips et al. (2008)]{phil08} Phillips, Kenneth J. H., Feldman, U., \& Landi, E. Ultraviolet and X-ray Spectroscopy of the Solar Atmosphere, by Kenneth J. H. Phillips; Uri Feldman and Enrico Landi. ISBN 978-0-521-84160-3 (HB). Published by Cambridge University Press, Cambridge, UK, 2008
\bibitem[Pierrard \& Lazar (2010)]{pier10} Pierrard, V. \& Lazar, M. 2010, SoPh, 267, 153
\bibitem[Pollock (2017)]{poll17} Pollock, C. J., Burch, J. L., Chasapis, A. et al. 2017, AGUFMSM, 11, 07
\bibitem[Raymond et al. (2010)]{raym10} Raymond, J. C., Winkler, P. F., Blair, W. P. et al. 2010, \apj, 712, 901
\bibitem[Smith(2001)]{smit01} Smith, R. K., Brickhouse, N. S., Liedahl, D. A., \& Raymond, J. C. 2001, \apj,
556, L91
\bibitem[Smith \& Hughes(2010)]{smit10} Smith, R. K., \& Hughes, J. P. 2010, \apj, 718, 583
\bibitem[Stepney \& Guilbert (1983)]{step83} Stepney, S. \& Guilbert, P. W. 1983, \mnras, 204, 1269
\bibitem[Suleimanov et al.(2005)]{sule05} Suleimanov, V., Revnivtsev, M., \& Ritter, H. 2005, A\&A, 435, 191
\bibitem[Summers \& Thorne (1991)]{summers83} Summers, D. \& Thorne, R. M, 1991, PhFlB, 3, 1835
\bibitem[Summers et al.(2006)]{summ06} Summers, H. P., Dickson, W. J., O¡¯Mullane, M. G., et al. 2006, Plasma Phys.
Control. Fusion, 48, 263
\bibitem[Sutherland(1998)]{suth98} Sutherland, R. S. 1998, \mnras, 300, 321
\bibitem[Svensson (1982)]{Svensson82} Svensson, R. 1982, \apj, 258, 335
\bibitem[Testa et al.(2014)]{test14} Testa, P., De Pontieu, B. Allred, J., et al. 2014, Sci., 346, 315
\bibitem[van Hoof et al.(2014)]{van14} van Hoof, P. A. M., Williams, R. J. R., Volk, K. et al. 2014, \mnras, 444, 420
\bibitem[van Hoof et al.(2015)]{van15} van Hoof, P. A. M., Williams, R. J. R., Volk, K. et al. 2015, \mnras, 449, 2112
\bibitem[Vasyliunas(1968)]{vasy68} Vasyliunas, V. M. 1968, in Physics of the Magnetosphere, eds. R. D. L.
\bibitem[Wannawichian et al.(2003)]{wann03} Wannawichian, S., Ruffolo, D., \& Kartavykh, Y.Y. 2003, \apjs, 146, 443
\bibitem[Yuasa et al.(2010)]{yuas10} Yuasa, T., Nakazawa, K., Makishima, K. et al. 2010, A\&A, 520, 25
\end{thebibliography}
\end{document}